\documentclass[aps,prl,twocolumn,amsmath,amssymb,amsbsy,superscriptaddress,floatfix,showpacs]{revtex4}
\def\up{\uparrow}
\def\down{\downarrow}
\def\eps{\varepsilon}
\def\ba{\begin{eqnarray}}
\def\ea{\end{eqnarray}}
\def\be{\begin{equation}}
\def\ee{\end{equation}}
\usepackage[dvips]{graphicx}
\begin{document}
\title{Non-analytic behavior of 2D itinerant ferromagnets.}
\author{Dmitry V. Efremov}
\affiliation{ Institute for Theoretical Physics, Technical
University of  Dresden, 01062 Dresden, Germany. }
\author{Joseph J. Betouras}
\affiliation{ Scottish Universities Physics Alliance, Department of
Physics and Astronomy,
 University of St. Andrews, St. Andrews KY16
9SS, Scotland, UK.}
\author{Andrey Chubukov}
\affiliation{ Department of Physics,
 University of Wisconsin-Madison, 1150 University Ave.,Madison, WI 53706, USA. }
\date{\today}
\begin{abstract}
We consider an ordered ferromagnet
in the vicinity of a $T=0$ transition into a paramagnet.
We show that the free energy and the
transverse and longitudinal static susceptibilities
contain non-analyticities which destroy
 a continuous second-order transition.
Depending on the parameters, the transition either becomes first-order, or
occurs via  an intermediate spiral phase.
\end{abstract}

 \maketitle

{\it Introduction.} In recent years, there has been a strong interest to
 understand hidden features of a
 $T=0$ ferromagnetic transition in itinerant fermionic systems. A much studied Hertz-Millis-Moriya (HMM) \cite{Hertz, Moriya, Millis} $\phi^4$ model of a ferromagnetic
quantum criticality predicts that the transition should be continuous in all dimensions $D>1$, with mean-field exponents,  like a classical
 transition in $D > 4=1+z$, where $z=3$ is the dynamical exponent.
 However, it was  realized \cite{Belitz1,Belitz2,Andrey1,Rech} that the difference between quantum and classical cases is more than the change of the effective dimension --
 in the quantum case, the $\phi^4$ and higher-order terms in the expansion in
 the order parameter field contain singular dynamic parts, which reflect the fact that fermions give rise to long-range, dynamical interaction between collective spin excitations in itinerant fermionic systems.
These extra dynamic terms have been analyzed on the paramagnetic
side of the transition, both in $D=3$ (Ref. \onlinecite{Belitz2}) and $D=2$ (Ref.  \onlinecite{Maslov}),
 and have been found to give rise to two competing effects: (i) the
 expansion of the free energy in the
magnetic field is non-analytic, and the non-analytic term favors a
preemptive first-order transition to a state with a finite magnetization,
 (ii) the static spin susceptibility $\chi (q)$ is non-analytic
in $q$ and becomes negative at some $q_0$, signaling another preemptive  instability, this time towards a spiral phase. Which instability of a paramagnet comes first depends on the interplay between the prefactors for the analytic $\Delta^4$ and $q^2$ terms, but in any event, a continuous second-order ferromagnetic transition point is internally unstable.

In this paper, we consider what happens when the system approaches a
 ferromagnetic quantum-critical point (QCP) from the ordered state.
We show that the free energy of a quantum ferromagnet is non-analytic in the order parameter field $\Delta$, and the non-analytic term
favors a first-order transition into a paramagnet at some $\Delta_c$. We analyze the static spin susceptibility, which in the ferromagnetic phase has non-equal longitudinal and transverse components, $\chi_{\parallel} (q)$ and $\chi_{\perp} (q)$, and  show that both  are non-analytic in momentum $q$
and both become negative  at a finite $q$ inside the ferromagnetic case,
 when  $\Delta$  becomes smaller that some critical
 $\Delta_{c\perp}$ and $\Delta_{c\parallel}$.  The negative $\chi_{\parallel,\perp} (q)$ implies the development of the spiral order, either
 along the direction of the magnetization, or in a transverse direction.
 We relate $\Delta_c, \Delta_{c\perp}$ and $\Delta_{c\parallel}$ with the parameters of the model and argue that if $\Delta_c > \Delta_{c\perp}, \Delta_{c\parallel}$, the transition is first order, otherwise the ferromagnetic phase first becomes unstable against a spiral.

Our analysis is based on the Eliashberg-type consideration near a
ferromagnetic  QCP. Such approximation has been justified in ~
\cite{Andrey1,Rech,Maslov} and we assume that it is valid. For
definiteness, below we only consider the case $D=2$, where the
non-analyticities are stronger than in $D=3$.

The point of departure of our analysis is the spin-fermion model
near a QCP. It describes fermions interacting with their own
collective excitations in the spin channel, described by spin
variables ${\bf S} = c^\dagger_{\alpha} {\bf \sigma}_{\alpha\beta}c_\beta$.
 The model does not
assume  a long-range order a 'priori,
 and is described by the Hamiltonian with three terms:
 $H_{f} = \sum_{k, \alpha=\up,\down} \varepsilon_{\mathbf{k}} c^{\dagger}_{\mathbf{k}, \alpha}c_{\mathbf{k}, \alpha}$,
 which describes low-energy fermions with the dispersion
$\varepsilon_{\mathbf{k}} \approx \varepsilon_F + v_F (k-k_F)$,
 $H_s = \sum_{q} \chi^{-1}_0 (q) \mathbf{S}_{\mathbf{q}} \mathbf{S}_{-\mathbf{q}}$, which describes
 collective bosonic excitations with a bare  static propagator
$\chi_0 (q)$,
 and the spin-fermion interaction term  $H_{int} = g~ (1/N)\sum_{\mathbf{k},\mathbf{q}, \alpha, \alpha'} c^{\dagger}_{\mathbf{k}, \alpha} \mathbf{\sigma}_{\alpha, \alpha'}
\mathbf{S}_{\mathbf{q}}c_{\mathbf{k}+\mathbf{q} , \alpha'}$.
Within this model, $\chi (q, \Omega_m) = \chi_0 (q)/(1- 2 g^2 \chi_0
(q) \Pi (q, \Omega_m))$ where $\chi_0 (q)$ is the static
susceptibility of free fermions, and $\Pi (q, \Omega_m)$ is the
polarization operator (here and below we set
 the Bohr magneton $\mu_B =1$).
Near a ferromagnetic transition,  $\chi (q) = 2 \nu(\varepsilon_F)/
(\delta + (aq)^2)$, where $\nu(\varepsilon_F)$ is the density of
states per particle  at the Fermi surface ($\nu(\varepsilon_F) =
m/(2\pi)$ for $\varepsilon_0 (k) = k^2/(2m)$),  $\delta = 1 - (2g
\nu(\varepsilon_F))^2$, and the length  $a$ is proportional to
the radius of the interaction~\cite{Maslov}.  The
 dynamic fermionic self-energy and the
dynamic part of $\Pi (q, \Omega_m)$
 are computed self-consistently within the model. The fully renormalized spin susceptibility in the paramagnetic
 phase is
\be \chi_{ij} (q, \Omega_m) = \delta_{ij}~ \frac{2
\nu(\varepsilon_F)}{\delta + (aq)^2 + {\bar \Pi} (q, \Omega_m)}, \label{4}
\ee
where ${\bar \Pi} (q, \Omega_m) = \Pi (q, \Omega_m) - \Pi (q, 0)$.
For free fermions
 ${\bar \Pi} (q, \Omega_m)= |\Omega_m|/\sqrt{(v_F q)^2 + \Omega^2_m}$.

The ferromagnetic transition occurs when $g = 1/(2
\nu(\varepsilon_F))$.  At larger $g$, $\delta$ becomes negative, and
 the system develops  a ferromagnetic long-range order with the magnetization $M =\langle S^z \rangle =  (N_\up -N_{\down})$. We
 assume that such long-range order does exist and search for preemptive
instabilities upon approaching QCP from the ferromagnetic side.

{\it Mean-field analysis}~~~  At the mean-field level, the
interaction term reduces to $ g \langle S_z \rangle \sum_{k,\alpha}
 \mbox{sign} \alpha~ c^{\dagger}_{\mathbf{k}, \alpha}c_{\mathbf{k}, \alpha}$, and the fermionic propagator becomes
\be
 G^{-1}_{\up,\down}(k, \omega_m) = i\omega_m -(\varepsilon_k - \mu)
  \pm \Delta_0,
\label{0} \ee where $\Delta_0 = g \langle S_z \rangle = g (N_{\up} -
N_{\down})$, and  $\mu = \mu (\Delta_0)$ is the exact chemical potential
(to order $\Delta^2_0$, $\mu (\Delta_0) = \mu (0) - \nu'/(2\nu)
\Delta_0^2 $, where $\nu'$ is the derivative of the density of
states at the Fermi surface). Because $\Delta_0$ is finite, the
longitudinal and transverse spin propagators become unequal already
in the static limit. We have
\ba
&&\chi^{zz}(q) = \frac{\chi^{zz}_{0}}{1 - (g \chi^{zz}_{0})^2 +
(aq)^2}, \nonumber \\
&&\chi^{xx}(q) = \frac{\chi^{xx}_{ 0}}{1 - (g \chi^{xx}_ {0})^2 + (aq)^2},
 \label{s1} \ea
where
\ba
&& \chi^{zz}_{0} = -\int \frac{d\omega_m}{2\pi} \int \nu(\varepsilon)
d \varepsilon \left(G^2_{\up} (\varepsilon, \omega_m) + G^2_{\down}
(\varepsilon, \omega_m)\right), \nonumber \\
&&
\chi^{xx}_{0} = -2 \int \frac{d\omega_m}{2\pi} \int \nu(\varepsilon)
d \varepsilon G_{\up} (\varepsilon, \omega_m)G_{\down} (\varepsilon, \omega_m).
\ea
A simple calculation shows that
$\chi^{xx} (0) = (N_{\up} - N_{\down})/\Delta_0 = 1/g$, such that
$\chi^{xx} (q)$ diverges at $q \rightarrow 0$, in agreement with the
Goldstone theorem. For longitudinal susceptibility, $\chi^{zz}(q) =
2 \nu (\varepsilon_F)/(\delta_F + (aq)^2)$, where
$\delta_F = -2 \delta = K \Delta^2_0$, and
$K = -(2/3)\nu''/\nu$ (Ref.\onlinecite{comm}).  The
 theory is only valid when $K >0$, otherwise the transition is first order by trivial reasons. By order of magnitude, $\delta_F
\sim (\Delta_0/\varepsilon_F)^2$.

The dynamic terms  ${\bar \Pi}^{xx} (q, \Omega_m)$ and ${\bar \Pi}^{zz} (q, \Omega_m)$
also differ at $\Delta_0 \neq 0$. Evaluating them using fermionic
propagators from (\ref{0}), we obtain
\ba
&& {\bar \Pi}^{zz}(q, \Omega_m) =
\frac{| \Omega_m|}{\sqrt{(v_F q)^2+\Omega^2_m}}, \nonumber \\
&&{\bar \Pi}^{xx}(q,i\Omega_m) = \frac{| \Omega_m|}{\sqrt{(v_F q)^2+(\Omega_m+ 2i\Delta_0)^2}}. \label{2}
\ea
The mean-field dynamic spin susceptibilities in the ferromagnetic phase are then given by
\ba
&&\chi^{zz}(q,\Omega_m)  =   \frac{2 \nu (\varepsilon_F)}{\delta_F + (aq)^2 + {\bar \Pi}^{zz}(q,\Omega_m)},
\nonumber \\
&&\chi^{xx} (q,i\Omega_m) =   \frac{2 \nu (\varepsilon_F)}
{(aq)^2 + {\bar \Pi}^{xx}(q,\Omega_m)}.
 \label{2a}
 \ea

{\it Eliashberg theory}~~
Eqs. (\ref{0}-\ref{2a}) constitute the mean-field description
 of the ferromagnetic  phase. Within this description, the transition is continuous, i.e., the ferromagnetic phase is stable up to a point where $\Delta_0, \delta =0$.  The Eliashberg theory goes beyond this approximation --
 it self-consistently takes into account  $\omega$-dependent
fermionic self-energy, but neglects $k-$dependent self-energy and vertex corrections. Vertex corrections generally are not small if the interaction involve small momentum transfers, and  are necessary to satisfy Ward identities related to the conservation laws. However, the analysis of
vertex corrections on the paramagnetic side
 have shown~\cite{Rech,Maslov} that they can be rigorously
neglected in the calculations of the non-analytic terms in the free energy and spin susceptibilities, if the interaction is sufficiently long-ranged such that $a k_F > 1$, which we assume to hold~\cite{comm_3}.

The calculations  proceed in three steps, like in the paramagnetic phase.
 First, we obtain self-consistent
 one-loop expressions for the fermionic self-energy and dynamic spin susceptibilities. Second, we use these one-loop expressions as
 inputs, obtain the
free energy within Eliashberg theory,
 and show that it is non-analytic in $\Delta$. Third, using the same
 inputs, we compute static spin susceptibilities at the two-loop level and
find terms which are non-analytic in momentum.
 We argue that the non-analytic terms in the free energy favor
a first-order transition, while the non-analytic terms
in the susceptibilities favor an intermediate spiral phase.

{\it Fermionic self-energy}~~~ The one-loop fermionic self-energy in the ordered phase is given by
\be
\Sigma_{f}(\omega_m) = \left\{
\begin{array}{lcl}
\lambda(\varepsilon_k) ~\omega_m & \mbox{~for~} & \omega_m \ll \omega_0/\lambda^3 \nonumber \\
\omega_0^{1/3} \omega_m^{2/3} & \mbox{~for~} &
 \omega_m \gg \omega_0/\lambda^3
\end{array}
\right.
\label{3}
\ee
where $\omega_0 =  3\sqrt{3} \varepsilon_F/(4 (ak_F)^4)$, $ \varepsilon_F =
v_F k_F/2$, and
$\lambda (\varepsilon_k)$ depends in non-singular way
on the ratio of $\varepsilon_k$ and $\Delta_0$. For $\varepsilon_k =\mu$
\be
\lambda = \frac{1}{4 ak_F} \left(\frac{v_F}{a \Delta_0} +
\frac{1}{\sqrt{\delta_F + (2 a \Delta_0/v_F)^2}}\right).
\ee
The self-energy is linear in $\omega_m$ at the smallest frequencies, but crosses over to the quantum-critical, $\omega_m^{2/3}$ behavior at frequencies larger than $\omega_0/\lambda^3 \sim \Delta_0^3/(\varepsilon^2_F (ak_F))$.
Such self-energy does not destroy the ferromagnetic order
 and preserves a Fermi surface, but it is larger than $\omega_m$ near QCP, and has non-Fermi liquid form in between $\omega_0/\lambda^3 \ll \omega_0$ and $\omega_0$.  The non-Fermi liquid behavior
 in the ferromagnetic state
has been discussed from a different point of view in Ref. \cite{Mineev}.

The one-loop dynamic polarization operators ${\bar \Pi}^{xx}_1 (q,\Omega_m)$ and ${\bar \Pi}^{zz}_1 (q,\Omega_m)$,  re-evaluated with dressed fermions,
 are given by rather complex expressions. Like in previous studies~\cite{Rech,Maslov}, we found  that, for the calculations of the non-analytic terms in the free energy and spin susceptibilities, we only need
 terms up to order $1/q^3$ in the $1/q$ expansion of ${\bar \Pi}(q,\Omega_m)$.
Such terms are not affected by vertex corrections~\cite{Rech}.
We obtained
\ba
&& {\bar \Pi}^{xx}_1 (q,\omega) = \frac{|\omega|}{v_F q}  \left(1 - \frac{ (\omega+ c_\omega \Sigma_{f}(\omega) + 2 i \Delta_0 )^2}{2 v^2_F q^2}\right)
, \nonumber \\
&& {\bar \Pi}^{zz}_1 (q,\omega) = \frac{|\omega|}{v_F q} \left(1 - \frac{(\omega+ c_\omega \Sigma_{f}(\omega))^2}{2 v^2_F q^2}\right).
\label{n1}
\ea
where $c_\omega$ interpolates between $c_{\omega =0}=1$ and $c_{\omega}
\approx 1.20$ for $\omega_0/\lambda^3 <\omega < \omega_0$.

{\it The free energy}~~~
The free energy per particle for the ferromagnetic
spin-fermion model in the Eliashberg approximation is given by
$$
\Xi = \Xi_0 (\Delta_0) + \Xi^{zz} + 2 \Xi^{xx},
$$
where
$\Xi_0 (\Delta) =  \frac{2m}{p^2_F} \left[ -(1/4) \delta_F
\Delta^2 +  (K/8)\Delta^4 + ...\right]$  is analytic in $\Delta$
Minimizing $\Xi_0$ and expanding around the minimum,
 we obtain the equilibrium
$\Delta = \Delta_0 = (\delta_F/K)^{1/2}$,
and reproduce the mean-field
 expression for the static $\chi^{zz} (q \rightarrow 0)$.
  Further,
\ba
&&\Xi^{zz}= \frac{V}{2N} \int \frac{d\Omega_m d^2q}{(2\pi)^3}
 \ln{\frac{\chi^{zz} (0,0)}{\chi^{zz} (q, \Omega_m)}} , \nonumber \\
&& \Xi^{xx}= \frac{V}{2N} \int \frac{d\Omega_m d^2q}{(2\pi)^3}
 \ln{\frac{\chi^{xx} (0,0)}{\chi^{xx} (q, \Omega_m)}},
\label{8}
\ea
where $\chi^{ii} (q, \Omega_m)$ include one-loop
polarization operators ${\bar \Pi}^{ii}_1 (q, \Omega_m)$.
Both $\Xi^{zz}$ and $\Xi^{xx}$ contain analytic contributions which
renormalize constants in $\Xi_0 (\Delta)$.
 These renormalizations are small in $1/(ak_F)$ (Ref.\cite{Maslov})
 and we neglect them.
 In addition,   $\Xi^{xx}$  contains the non-analytic term in
 $\Delta$, which is our primary interest.
Substituting  $\chi^{xx}$ with the polarization operator from
(\ref{n1}) into Eqn. (\ref{8}), integrating over momentum,
 and neglecting regular terms, we obtain at QCP
\be
 \Xi^{xx} = \frac{\sqrt{2}}{2\pi c^{3/2}}
~\frac{\Delta^{7/2}}{\varepsilon^2_F \omega_0^{1/2}} Z , \label{n2}
\ee where $Z$ is the universal (cutoff independent) part of the
integral
\be Z = 2 \int_{0}^A d y Re \left[(y^{2/3} -i)^2
\log{(y^{2/3} -i)^2}\right]. \label{n3_1} \ee
The evaluation of the integral yields
 $Z = -8\pi \sqrt{2}/35 \approx -1.02$. Combining
non-analytic and analytic terms, we obtain in the immediate vicinity
of the QCP
\be \Xi = \frac{1}{\varepsilon_F} \left[
 -\frac{\delta_F}{4}
\Delta^2 + \frac{K}{8} \Delta^4 - \frac{\Delta^{7/2}}{E^{3/2}} \right],
\label{n3}
\ee
where
 $E = d \varepsilon_F/(ak_F)^{4/3}$, and $d
 \approx 3.50$.  Apart for
 a small numerical difference in $d$, this
 expression coincides with the one obtained in~\cite{Maslov},
where QCP was approached from the paramagnetic side.
There is, however, an important distinction between paramagnetic and
ferromagnetic phases away from QCP. In the paramagnetic phase, the
$\Delta^{7/2}$ dependence
 is replaced by $\Delta^3$ when $\Delta/\varepsilon_F \sim
 \delta/(ak_F)^2 \leq \delta$.
In the ferromagnetic phase, $\delta \rightarrow \delta_F $, which
  by itself scales as $(\Delta/\varepsilon_F)^2$.
As a result, the non-analytic  $\Delta^{7/2}$ dependence of $\Xi$ survives
 in the wide range away from a QCP,
 as long as $\Delta/\varepsilon_F \leq  \gamma$, where  $\gamma =(a k_F)^2/(K \varepsilon_F^2)$.

\begin{figure}[tbp]
\centering
\includegraphics[width=0.7\columnwidth]{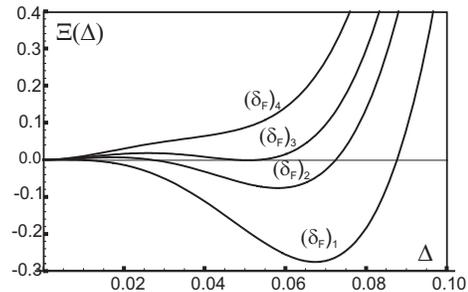}
\caption{The free energy $\Xi (\Delta)$, Eq. (\protect\ref{n3}),
for different values of $\delta_F$, going from positive $(\delta_F)_1$ to negative $(\delta_F)_{2,3,4}$. The first-order transition occurs
at $(\delta_F)_3$.}
\label{fig1}
\end{figure}

The negative $\Delta^{7/2}$ term in the free energy shifts the
 equilibrium value  $\Delta_0$ such that it
remains finite even on the paramagnetic side of the transition,
when $\delta_F$ changes sign and becomes negative.
For a generic $\delta_F$, $\Delta_0$ is the solution of
$K \Delta^2_0 - 7 (\Delta_0/E)^{3/2} = \delta_F$.
 One can easily verify (see Fig.\ref{fig1}) that the free
 energy (\ref{n3}) describes a first-order transition to a paramagnet at
 $\delta_F =- (\Delta_0/E)^{3/2} <0$ (i.e., $2\delta = -\delta_F >0$).
 The value of the equilibrium $\Delta_0$ at such transition is
$\Delta_c = 36 /(K^2 E^3) \approx 0.84 \gamma^2  \varepsilon_F$.
 We also note that the stiffness --
 a prefactor for $(\Delta -\Delta_0)^2/2$ in the free energy --
 changes from $\delta_F$ to $\delta^{eff}_F =
 \delta_F + (7/4) (\Delta_0/E)^{3/2}$, and  remains  positive
for all $\Delta_0 > \Delta_c$.  At  $\Delta_0 =
\Delta_c$, $\delta^{eff}_F = (3/4) |\delta_F|$. \\

{\it Static spin susceptibilities}~~~
We next show that the static spin susceptibilities $\chi^{xx} (q,0)$ and $\chi^{zz} (q,0)$ also display
 non-analytic behavior,  and that these
 non-analyticities compete with the one in the free energy and may give rise
 to pre-emptive spiral instabilities.
The non-analytic term in $\chi (q,0)$
 has been previously analyzed on the paramagnetic side~\cite{Rech,Maslov}.
 We performed the  calculations in the ordered phase.

The non-analytic behavior of $\chi (q,0)$ originates from
non-analytic $q-$dependencies of $\Pi^{zz}$ and $\Pi^{xx}$, which
acquire static parts at the two-loop  order.

 The computational steps are similar to those in Ref.\cite{Rech} and we refrain from discussing them.
 The non-analytic contributions $\Pi^{zz}_2$ and $\Pi^{xx}_2$ come
 from the processes in which fermions and spin fluctuations
 are vibrating near a fermionic mass shell and are far away from a bosonic mass shell~\cite{Andrey1,Rech} (
 the same processes lead to fermionic self-energy $\Sigma_f (\omega_m)$).
We found that  the non-analyticity
 comes from the exchange processes involving  transverse spin fluctuations.

 There are two types of non-analyticities
 in an ordered ferromagnet. First, there are corrections to static, uniform
$\Pi^{zz} (0,0)$. They change $\delta_F$ into $\delta^{eff}_F$, which
 is the same as the stiffness obtained by expanding the free energy.
Second, there are non-analytic terms in the momentum
expansion of the static $\Pi^{zz} (q, 0)$ and $\Pi^{xx} (q,0)$
For the latter, we found
\ba
&&\Pi^{zz}_2 (q,0) = - a^2 q^{3/2} k_F^{1/2} F^{zz}(2\Delta_0/v_F q),~~\nonumber \\
&& \Pi^{xx}_2 (q,0) = - a^2 q^{3/2} k_F^{1/2} F^{xx}(2\Delta_0/v_F q),
\label{5}
\ea
such that
\ba
&&\chi^{zz}(q,0) = \frac{2\nu(\varepsilon_F)}{a^2}~\frac{1}{\frac{\delta^{eff}_F}{a^2} +q^2 - q^{3/2} k_F^{1/2}  F^{zz}
\left(\frac{2\Delta_0}{v_F q}\right)}, \nonumber \\
&&\chi^{xx}(q,0) = \frac{2 \nu(\varepsilon_F)}{a^2}~\frac{1}{q^2 - q^{3/2} k_F^{1/2}  F^{xx}
\left(\frac{2\Delta_0}{v_F q}\right)}.
\label{6}
\ea
For $\Delta_0 \ll v_F q $,
 $F^{zz}(0) = F^{xx}(0) \approx 0.25$, in agreement with~
\cite{Rech}. For
  $v_F q \ll \Delta_0$  the two scaling functions differ.
The scaling function $F^{zz}(y)$  remains close to $F^{zz} (0)$ as long as the argument
$y < \varepsilon_F/\Delta_0$. The function
$ F^{xx}(y)$ crosses over to  $F^{xx}(y \gg 1) \approx 0.15/\sqrt{y}$
such that $ \Pi^{xx}_2 (q,0)$ becomes analytic, $\Pi^{xx}_2 (q,0) \propto q^2$.
\begin{figure}[tbp]
\centering
\includegraphics[width=0.85\columnwidth]{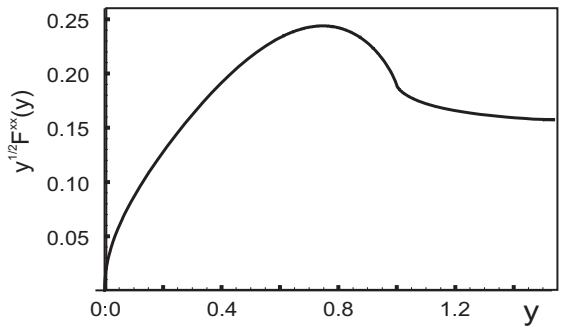}
\caption{The scaling function $\sqrt{y} F^{xx}(y)$.
It has a maximum value of $\sim 0.24$ at $y \approx 0.74$}
\label{fig2}
\end{figure}
The function $\sqrt{y} F^{xx} (y)$ is plotted in Fig. 2
Analyzing Eq. (\ref{6}), we find that both susceptibilities become
negative at a finite $q$ when  $\Delta_0$ reduces below some critical value.
The transverse susceptibility
 becomes negative at  $\Delta_0  = \Delta_{c,\perp} =  \varepsilon_F (\sqrt{y} F(y))^2_{max} = 0.06 \varepsilon_F$,
 whereas  longitudinal susceptibility
 becomes negative at $\Delta_0 = \Delta_{c,\parallel}$
, which is the
solution of
$$\frac{\Delta_{c,\parallel}}{\eps_F} \left(1 - \frac{21}{4} \frac{\gamma}{d^{3/2}} \sqrt{\frac{\eps_F}{\Delta_{c, \parallel}}} \right)^{1/2}
 = 0.02  \sqrt{\gamma}.$$

\begin{figure}[tp]
\centering
\includegraphics[width=0.75\columnwidth]{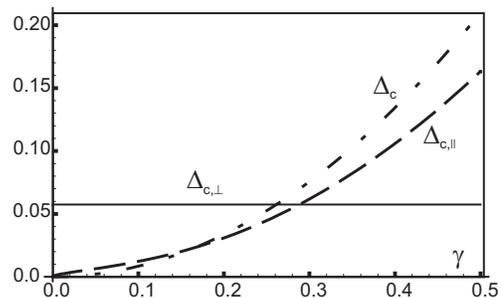}
\caption{The three critical  $\Delta_{c}$ as functions of  the
 parameter $\gamma = (a k_F)^2/(K \varepsilon_F^2)$. }
\label{fig3}
\end{figure}

 Whether the transition is the first order or involves an intermediate spiral phase depends on which of
$\Delta_c$, $\Delta_{c,\parallel}$ and $\Delta_{c,\perp}$ is the
largest. All three critical $\Delta$ scale with $\varepsilon_F$, but
they depend differently on the parameter $\gamma$. In Fig.\ref{fig3}
we plotted the three critical $\Delta$ vs $\gamma$. We see that for
$\gamma<0.26$,~
 $\Delta_{c,\perp} > \Delta_{c,\parallel},~\Delta_{c}$, and
   the system first develops a transverse spiral order, while
for  $\gamma>0.26$,
 $\Delta_c > \Delta_{c,\parallel},~\Delta_{c,\perp}$,
 and the transition is first order.
 This qualitatively agrees with the analysis on the paramagnetic side~\cite{Maslov}.

In summary, we analyzed the non-analytic terms in the free energy
 and in static susceptibilities in an ordered itinerant
 ferromagnet.  We found that,
 because of these non-analyticities,
 the transition to a paramagnet is  either first order,
 or involves an intermediate spiral phase.

We thank  A. Andreev,  P. Fulde,  A. Green, A. Huxley, D. Maslov, A. Rosch,  and
Yu. Ovchinnikov for useful discussions. The research has been supported
by NSF DMR 0604406 (A. Ch.) and in part by NSF PHY05-51164 (J.B.). J.B. would
like to thank KITP, Santa Barbara for hospitality during the
 completion of this work.


\begin{thebibliography}{99}
\bibitem{comm} A somewhat different result for $K$ is obtained if
  one uses RPA for the  Hubbard model.
In this case, charge component of the
interaction also contributes to $K$, and $K = (\nu'/\nu)^2 - (2/3) \nu ''/\nu$.
\bibitem{comm_3}  An alternative  way to justify Eliashberg theory even if $a k_F \leq 1$ is to extend the model
 to  $M \gg 1$ spieces of fermions, which separately interact with a collective boson. Vertex corrections are then small in $1/M$.
\end{thebibliography}
\end{document}